\documentclass[twocolumn, tighten]{aastex631}
\usepackage[utf8]{inputenc}
\usepackage{natbib,graphics,
graphicx,amsmath,amssymb,threeparttable,xcolor}

\newcommand{\ff}{$f_{\rm F444W}$}
\newcommand{\fo}{$f_{\rm F150W}$}

\shortauthors{McKay et al.}

\begin{document}

\title{Comparing SCUBA-2 and ALMA Selections of Faint DSFGs in Abell 2744}

\correspondingauthor{Stephen McKay}
\email{sjmckay3@wisc.edu}

\author[0000-0003-4248-6128]{S.~J.~McKay}
\affiliation{Department of Physics, University of Wisconsin--Madison, 1150 University Avenue,
Madison, WI 53706, USA}

\author[0000-0002-3306-1606]{A.~J.~Barger}
\affiliation{Department of Astronomy, University of Wisconsin--Madison, 475 N. Charter Street, Madison, WI 53706, USA}
\affiliation{Department of Physics and Astronomy, University of Hawaii, 2505 Correa Road, Honolulu, HI 96822, USA}
\affiliation{Institute for Astronomy, University of Hawaii, 2680 Woodlawn Drive, Honolulu, HI 96822, USA}

\author[0000-0002-6319-1575]{L.~L.~Cowie}
\affiliation{Institute for Astronomy, University of Hawaii, 2680 Woodlawn Drive, Honolulu, HI 96822, USA}

\begin{abstract}
We make a comparison of deep SCUBA-2 450~$\mu$m and 850~$\mu$m imaging 
on the massive lensing cluster field Abell~2744 with Atacama Large
Millimeter/submillimeter Array (ALMA) 1.2~mm data.
Our primary goal is to assess how effective the wider-field SCUBA-2 sample, in combination with
red JWST priors, is for finding faint dusty star-forming galaxies (DSFGs) compared to the much more
expensive mosaicked ALMA observations. We cross-match our previously reported direct ($>5\sigma$)
SCUBA-2 sample and red JWST NIRCam prior-selected ($>3\sigma$) SCUBA-2 sample to direct 
ALMA sources from the DUALZ survey. We find that roughly 95\% are confirmed by ALMA. 
The red priors also allow us to probe deeper in the ALMA image. Next, by measuring
the 450~$\mu$m and 850~$\mu$m properties of the full ALMA sample, we show that
46/69 of the ALMA
sources are detected at 850~$\mu$m and 24/69 are detected at 450~$\mu$m in the SCUBA-2 images, with a total detection fraction of nearly 75\%. All of the robust ($>5\sigma$) ALMA
sources that are not detected in at least one SCUBA-2 band lie at 1.2~mm fluxes $\lesssim$~0.6~mJy and are undetected primarily
due to the higher SCUBA-2 flux limits. We also find that the SCUBA-2 detection fraction drops slightly beyond $z=3$, which we attribute to the increasing 1.2~mm to 850~$\mu$m and 1.2~mm to 450~$\mu$m flux ratios combined with the ALMA selection.
The results emphasize the power of combining SCUBA-2 data with JWST colors to map the faint DSFG population.
\end{abstract}

\section{Introduction}

The demonstration by the COBE satellite that the cosmic infrared background (CIB) is comparable to the cosmic optical background showed that a significant fraction of the UV/optical starlight in the Universe is absorbed and reradiated by dust \citep{puget96,fixsen98,hauser98,dole06}. Resolving the CIB into individual sources has been one of the fundamental goals of modern astronomy for over two decades, which led to the submillimeter detection of extremely luminous, dusty star-forming galaxies (DSFGs; or submillimeter galaxies, SMGs) with star formation rates (SFRs) up to $\sim$1000~M$_\odot$/yr \citep{smail97, barger1998, hughes1998, eales1999}. 

Due to the relatively low surface density of DSFGs and the strong negative K-correction at submillimeter and millimeter wavelengths, wide-field single-dish submillimeter/millimeter surveys have been most effective at obtaining large samples of these galaxies. Currently, the most efficient instruments for this are SCUBA-2 \citep{holland13} at 450~$\mu$m and 850~$\mu$m on the James Clerk Maxwell Telescope (JCMT), NIKA2 \citep{monfardini+14,calvo+16} at 1.2~mm and 2~mm on the IRAM 30~m telescope, and soon TolTEC \citep{wilson20} from 1.1~mm to 2~mm on the Large Millimeter Telescope (LMT). 

However, single-dish submillimeter surveys are typically limited to picking out bright SMGs ($f_{850\mu{\rm m}} > 2$~mJy), because the low angular resolution (e.g., 14$''$ for SCUBA-2 at 850~$\mu$m) causes the blending of faint unresolved sources to dominate the noise below a certain flux level (the confusion limit; e.g., $\sim$1.6~mJy for SCUBA-2 at 850~$\mu$m; \citealt{cowie17}). The bright SMG population constitutes $\sim20$--30\% of the CIB.
An effective method for pushing to deeper survey depths is to leverage the gravitational lensing of massive galaxy clusters \citep[e.g.,][]{smail97,cowie02}, which offers the additional benefit of increased spatial resolution.
Such studies show that the majority of the CIB is made up of faint SMGs ($f_{850\mu{\rm m}} \lesssim 1$~mJy) \citep[e.g.,][]{knudsen08,chen13,hsu16}. 

In the last decade, substantial efforts have been made to study the faint end of the DSFG population. Interferometers, such as the Atacama Large Millimeter/submillimeter Array (ALMA), are capable of probing to deeper flux limits than SCUBA-2, but the small field-of-view means that direct searches using mosaicked observations \citep[e.g.,][]{hatsukade16,aravena16,walter16, dunlop17,umehata17,gonzalez-lopez17,munoz-arancibia18, hatsukade18, gonzalez-lopez20, gomez-guijarro22, casey21, zavala21, munozarancibia+23, fujimoto+23a,fujimoto+23b} are very costly. 

Using JWST data from multiple NIRCam programs (JWST-ERS-1324, JWST-GO-2561, JWST-DDT-2756)
on the Abell 2744 (A2744) cluster field \citep{paris23,weaver23}, 
along with ALMA data from the ALMA Lensing Cluster Survey (ALCS; 1.2~mm) \citep{fujimoto+23a} and the ALMA Frontier Fields Survey (AFFS; 1.1~mm) \citep{gonzalez-lopez17, munozarancibia+23} and SCUBA-2 data from \citet{cowie22},
\citet{barger23} developed a JWST NIRCam red galaxy selection that identified all known ($> 4.5\sigma$) ALMA sources (9) and nearly all known ($>5\sigma$) SCUBA-2 850~$\mu$m sources (17 out of 19). They then used the red galaxy priors to probe deeper into the SCUBA-2 data ($>3\sigma$), finding sources
down to faint intrinsic 850~$\mu$m fluxes of $\sim$0.1~mJy with the help of gravitational lensing. 

\citet{fujimoto+23b} subsequently presented the DUALZ ALMA 1.2~mm  survey, which maps most of the JWST NIRCam region of the A2744 UNCOVER program \citep{bezanson+22}. They released a catalog of 69 ALMA sources down to a minimum S/N of 4.2.
However, they did not cross-reference their sources with the SCUBA-2 data to see how many of their sources had already been detected. In addition to SCUBA-2 covering much larger areas than ALMA, its extended wavelength coverage is important for studying the physical properties of DSFGs, since 450~$\mu$m and $850~\mu$m lie closer to the far-infrared (FIR) peak of the dust emission and thus can help constrain dust temperatures and redshifts \citep[e.g.,][]{casey13, zavala18, lim20, barger22}. 

In this paper, we revisit the overlap between the SCUBA-2 and ALMA samples in A2744, incorporating the new DUALZ data. 
In Section~2, we describe the data that we use in our analysis. In Section~3, we compare the ALMA sample to the direct and red galaxy prior-based SCUBA-2 samples, and we also recover fainter ALMA sources using the priors. In Section~4, we use the 450~$\mu$m and 850~$\mu$m data to learn more about the properties of the ALMA sources. In Section~5, we contrast the properties of sources detected in the SCUBA-2 maps against those that are not, and we assess how effective the red galaxy priors are for recovering SCUBA-2 and ALMA sources. Finally, in Section~6, we summarize our results.

We assume a flat concordance $\Lambda$CDM cosmology with $\Omega_M= 0.3$, $\Omega_\Lambda = 0.7$, and $H_0 = 70.0$~km~s$^{-1}$~Mpc$^{-1}$.

\begin{figure*}[!t]
    \centering
    \includegraphics[width=0.95\linewidth]{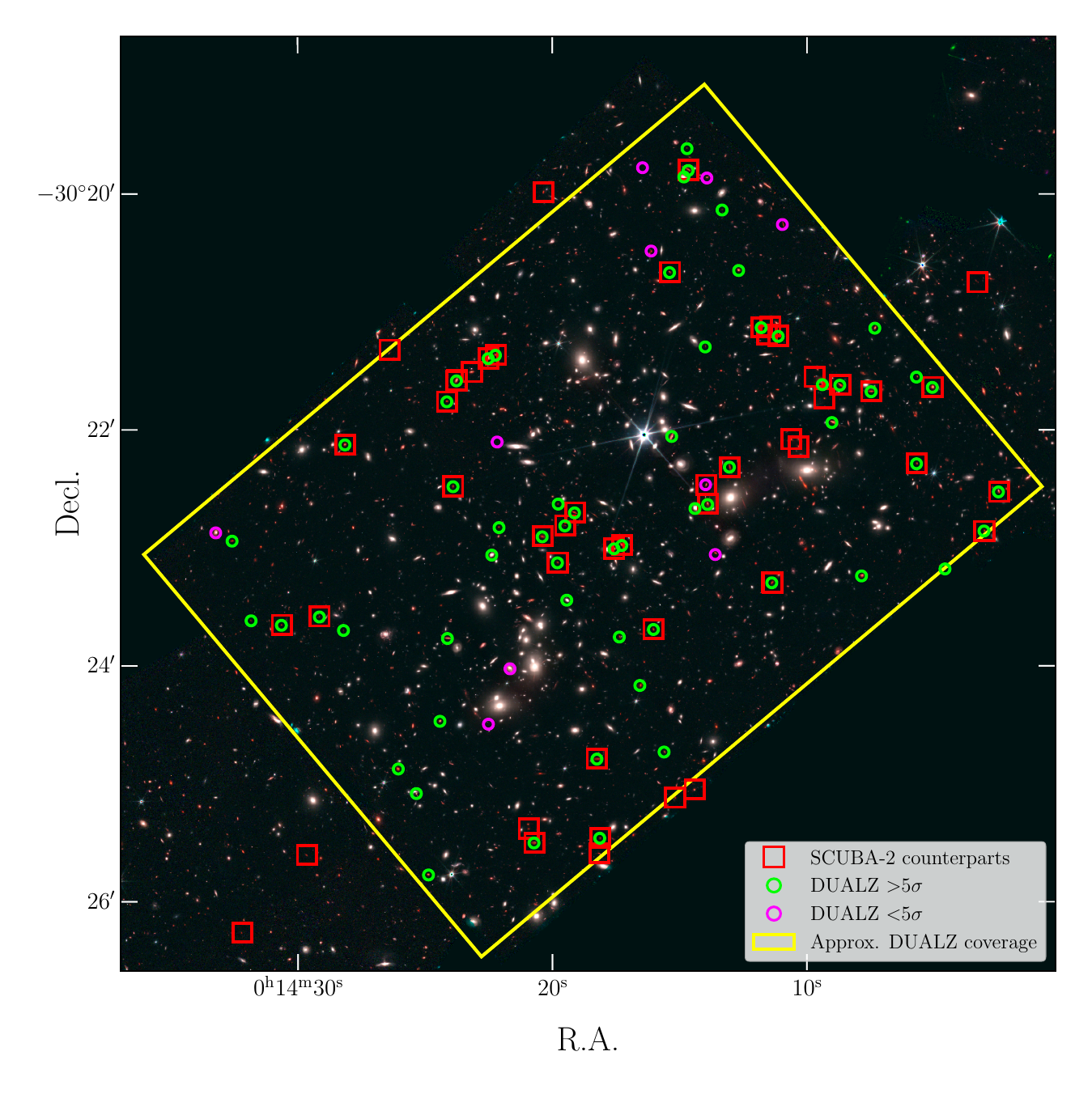}
    \caption{ JWST NIRCam RGB image (R: F444W, G: F150W, B: F115W) from \citet{paris23} with the outline of the approximate ALMA DUALZ coverage ($\sim$24~arcmin$^2$) from \citet{fujimoto+23b} overlaid in yellow. Red squares mark the positions of the SCUBA-2 sources with red NIRCam counterparts from \cite{barger23} (either direct $>5\sigma$ 850~$\mu$m detections or selected using the red galaxy priors; 49 sources). Note that one of the direct detections has a different red NIRCam counterpart confirmed by ALMA than was originally found by \citet{barger23}.
    We show the DUALZ sources in green if $>5\sigma$ (59 sources) and in magenta if $<5\sigma$ (10 sources). The SCUBA-2 imaging covers the entire NIRCam region and extends significantly beyond it. 
    }
    \label{fig:mosaic}
\end{figure*}

\begin{figure*}[ht]
    \centering
    \includegraphics[width=0.9\linewidth]{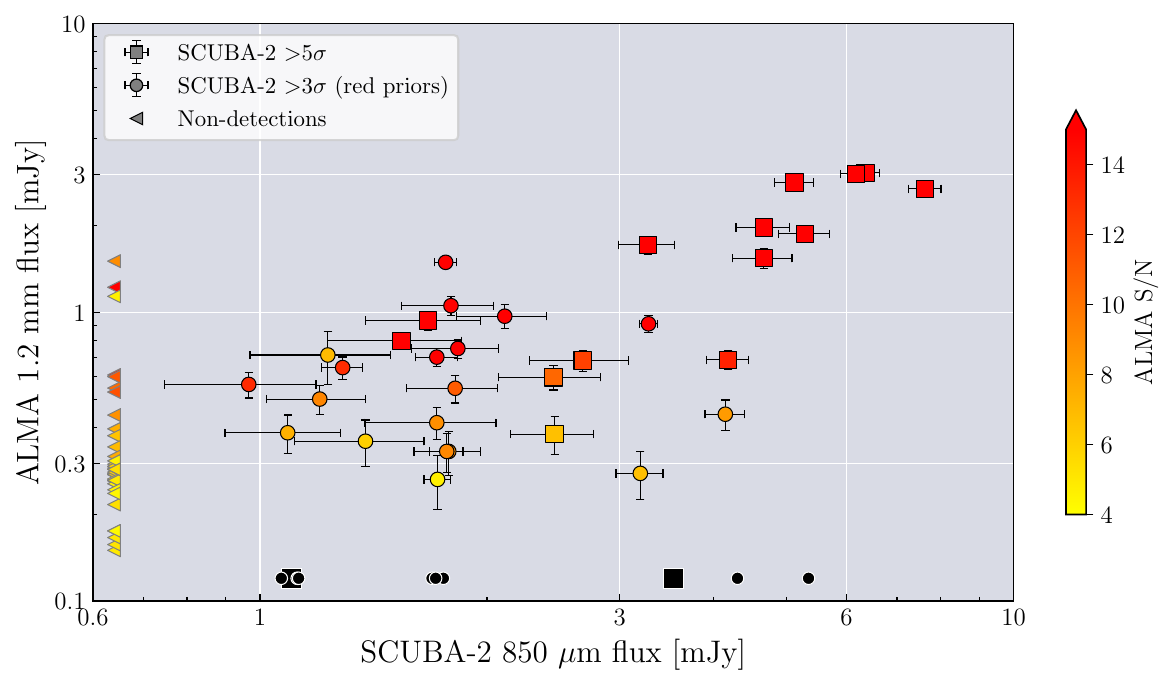}
    \caption{ALMA 1.2~mm flux density vs. SCUBA-2 850~$\mu$m flux density for sources in either the DUALZ or SCUBA-2 catalogs. The $>5\sigma$ SCUBA-2 sources are shown as squares. The $>3\sigma$ SCUBA-2 sources obtained with red NIRCam priors are shown as circles. The sources that are only in the DUALZ catalog are shown as leftward-facing arrows. The color of the points denotes the S/N of the ALMA detection using the righthand scale (black if not in the DUALZ catalog---note that there are four overlapping SCUBA-2 sources at $\sim$1.1~mJy on the figure).}
    \label{fig:1mm_850um}
\end{figure*}

\section{Data}

The massive lensing cluster field A2744 ($z=0.308$) is one of the six Hubble Frontier Fields \citep{lotz+17} and has been the target of hundreds of hours of multiwavelength observations from the UV to the radio. We have been observing A2744 with SCUBA-2 for a number of years and have generated deep 450~$\mu$m and 850~$\mu$m images covering the entire central region of the cluster \citep{cowie22}. The minimum 1$\sigma$ rms of the current 850~$\mu$m image is 0.26~mJy, while the minimum 1$\sigma$ rms of the current 450~$\mu$m image is 2.8~mJy, slightly deeper than those presented in \citet{cowie22}. The reduction of the SCUBA-2 maps and the source detection follow that described in \citet{chen13} and \citet{cowie17}.

A2744 has been observed in several JWST surveys, including deep NIRCam imaging from the UNCOVER \citep{bezanson+22}, GLASS-JWST \citep{treu22}, and DDT~\#2756 programs (PI: W. Chen; \citealt{roberts-borsani23}). In addition to lensing models released by five independent teams prior to JWST, \citet{furtak+23} published a lensing model for the cluster region using the UNCOVER JWST observations.

The ALMA data for this field come primarily from the DUALZ program \citep{fujimoto+23b}, which mapped the UNCOVER NIRCam area with ALMA Band 6 observations (effective wavelength of 1.2~mm) down to a minimum rms noise of 32.7 $\mu$Jy. The data from the 1.2~mm ALCS \citep{fujimoto+23a} and 1.1~mm AFFS \citep{gonzalez-lopez17, munozarancibia+23} surveys were combined into the \textit{Deep} maps. The \textit{Wide} maps use only the DUALZ observations. From the \textit{Deep} and \textit{Wide} maps (both with and without $uv$-tapering applied), \citet{fujimoto+23b} produced a source catalog of 69 ALMA sources down to a minimum S/N of 4.2 (their detection threshold varied depending on which map was used for the detection). In Figure~\ref{fig:mosaic}, we show their 69 sources and the coverage of the DUALZ survey overlaid on the JWST NIRCam image of \citet{paris23}.

\section{ALMA properties of SCUBA-2 sources}
\label{sec:alma_props}

Restricting to the DUALZ ALMA area,
we first compare the 17 $>5\sigma$ SCUBA-2 sources with the DUALZ catalog.
 \citet{barger23} found red NIRCam counterparts
(defined as \ff/\fo~$>$ 3.5 and \ff$ > 1$~$\mu$Jy) within 4$''$ of 15 of these SCUBA-2 sources
(note that in one case, two SCUBA-2 sources share one counterpart). 
The ALMA data confirm all but one of these;
for the remaining source, the ALMA detection matches to a different but similarly red galaxy at a slightly larger separation from the SCUBA-2 centroid. The two SCUBA-2 sources without a red NIRCam counterpart do not appear to have ALMA detections, though for one, there is an ALMA source at 5.3$''$ from the SCUBA-2 position. In summary, 15/17 (88\%) of the $>5\sigma$ SCUBA-2 detections have ALMA 1.2~mm counterparts.

With the ALMA data, we can check whether multiple DSFGs may be contributing to the 850~$\mu$m flux. We find that only two of the 15 $>5\sigma$ SCUBA-2 sources (SCUBA7 and SCUBA8) have multiple ALMA counterparts within $4\farcs5$ of the SCUBA-2 850~$\mu$m position, giving a multiplicity of 13\% (2/15).

\citet{barger23} also used the red NIRCam galaxies in the field as priors to identify fainter sources than could be securely extracted from the SCUBA-2 data alone. They compiled a sample of 44 priors with associated $>3\sigma$ SCUBA-2 fluxes (their Table~3), 39 of which fall within the DUALZ area. Twelve of these correspond to 13 of the 15 direct $>5\sigma$ SCUBA-2 sources with red NIRCam counterparts in the area, since two SCUBA-2 sources have the same red counterpart. 
The other two sources (SCUBA9 and SCUBA12) are not included in this sample due to differences in the SCUBA-2 deblending when using priors. We show the positions of all the red NIRCam counterparts to the SCUBA-2 sources in Figure~\ref{fig:mosaic} with red squares.

Thirty-one of the 39 sources match to the DUALZ catalog, while eight do not. Combining these 31 with SCUBA9 and SCUBA12,
the SCUBA-2 sources account for 33/69 (48\%) of the DUALZ ALMA sample. 
If we consider only the 59 $>5\sigma$ DUALZ sources, then the SCUBA-2 sources account for 32/59 (54\%). 

In Figure~\ref{fig:1mm_850um}, we show the 1.2~mm and 850~$\mu$m fluxes for any ALMA or SCUBA-2 source in the DUALZ area, distinguishing between those with direct $>5\sigma$ SCUBA-2 detections (squares) and those based on the red galaxy priors (circles). 
We show the ALMA sources not matched to a SCUBA-2 source as leftward-facing arrows, 
and the $>5\sigma$ SCUBA-2 detections not in the DUALZ catalog in black along the bottom. 
For the sources detected in both ALMA and SCUBA-2, the median 850~$\mu$m to 1.2~mm flux ratio is 2.6, with a 16--84th percentile range of 1.8 to 5.4. From Figure~\ref{fig:1mm_850um}, we can also see that most of the sources detected by DUALZ but not by SCUBA-2 are faint and/or lower signal-to-noise (S/N) ALMA sources (righthand color scale). 

Next, we investigate whether any of the eight $>3\sigma$ red prior-selected SCUBA-2 sources 
not in the DUALZ catalog are detected in the ALMA image at a lower significance. To test this, we measure the ALMA 1.2~mm fluxes at the positions of these eight SCUBA-2 $>3\sigma$ sources by finding the largest local peak within $1\farcs0$ of the red NIRCam prior, using the natural weighted DUALZ {\it Wide} map with primary beam correction applied. The local noise is estimated by measuring the dispersion in random pixels within 12$''$ of the source, after masking out any $>4\sigma$ sources nearby. 

\begin{figure}[!t]
    \centering
    \includegraphics[width=\linewidth]{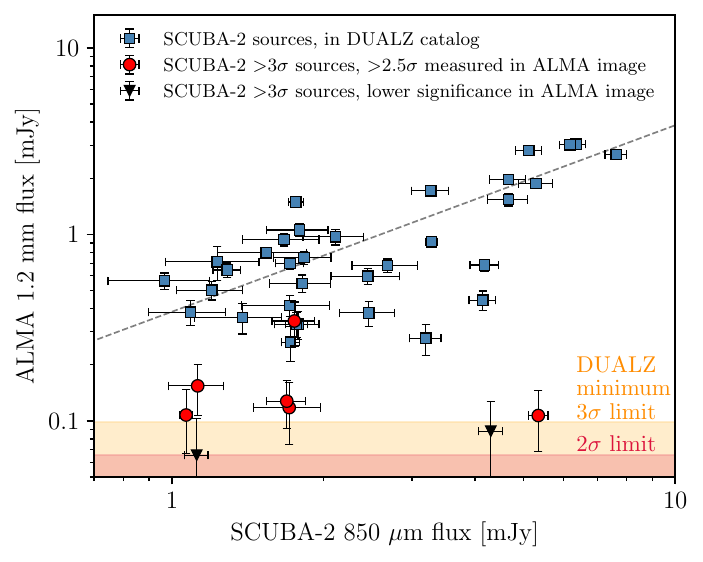}
    \caption{ALMA 1.2~mm vs. SCUBA-2 850~$\mu$m flux densities for the combined $>5\sigma$ and $>3\sigma$ SCUBA-2 sources (\citealt{barger23} Tables 2 and 3). Direct detections in the DUALZ catalog are shown as blue squares, while sources for which we measure the ALMA flux densities are shown as red circles if $>2.5\sigma$ and as black triangles if $<2.5\sigma$ in the DUALZ {\it Wide} map. Also shown are the minimum 3$\sigma$ and 2$\sigma$ limits (shaded regions; the actual limits may be higher than these at any given point). We show the median 850~$\mu$m to 1.2~mm ratio of 2.6 for the sources that are detected in the DUALZ catalog (dashed line). }
    \label{fig:measured_alma}
\end{figure}

We find that of these eight sources, six are detected at $>2.5\sigma$ (three at $>3\sigma$). However, we note that one of the $>3\sigma$ sources falls near the edge where the noise is higher, so the error measured by our method may be an underestimate of the actual uncertainty for this source. In Figure~\ref{fig:measured_alma}, we show the 1.2~mm and 850~$\mu$m fluxes for all the SCUBA-2 sources in the DUALZ area, including the eight sources for which we measure the ALMA fluxes. We show the SCUBA-2 sources that match to direct DUALZ detections as blue squares, and mark the median 850~$\mu$m to 1.2~mm flux ratio of 2.6 on the plot (dashed line). The six red prior-selected sources for which we measure $>2.5\sigma$ 1.2~mm fluxes are shown as red circles, while the two sources at lower significance are shown as black triangles. The deepest 2$\sigma$ and 3$\sigma$ limits of the DUALZ {\it Wide} map are also shown (shaded region), though the local 2$\sigma$ and 3$\sigma$ limits will generally be higher than this at any given point in the maps. This figure illustrates that the red NIRCam priors enable us to probe to fainter ALMA fluxes than would be possible with a direct search.

In total, we find that within the DUALZ area, 37/39 of the red prior-selected SCUBA-2 sources from \citet{barger23} are detected at $>2.5\sigma$ in the ALMA  image  (34/39 if restricted to $>3\sigma$), confirming the robustness of this selection. We summarize the numbers of direct $>5\sigma$ and prior-selected $>3\sigma$ SCUBA-2 sources and the fractions that have red galaxy and/or ALMA counterparts in Table~\ref{tab:samples}.

\section{SCUBA-2 450 and 850 micron properties of ALMA sources}

Having shown that a high fraction of the SCUBA-2 sources detected either directly or with the red galaxy priors are confirmed with ALMA, we next invert the analysis and study the SCUBA-2 450~$\mu$m and 850~$\mu$m properties of the 69 ALMA sources in the DUALZ catalog.  
All of these sources fall on the deep SCUBA-2 450~$\mu$m and 850~$\mu$m maps from \citet{cowie22}.

We extract the 450~$\mu$m and 850~$\mu$m fluxes and errors for the ALMA priors using an iterative procedure to avoid contamination of fainter sources by brighter ones: We first measure the SCUBA-2 flux at the position of each ALMA source; then, working down the catalog from brighter to fainter SCUBA-2 flux, we measure the peak flux and the error within 4$''$ from the ALMA position and subtract from the residual image the SCUBA-2 PSF scaled to the peak flux. We then repeat for the next brightest SCUBA-2 flux until all sources have been measured. Although this process is not strictly necessary for the 450~$\mu$m maps due to the higher angular resolution, we perform the iteration on both SCUBA-2 maps anyway. 

\begin{figure}[!t]
    \centering
    \includegraphics[width=0.99\linewidth]{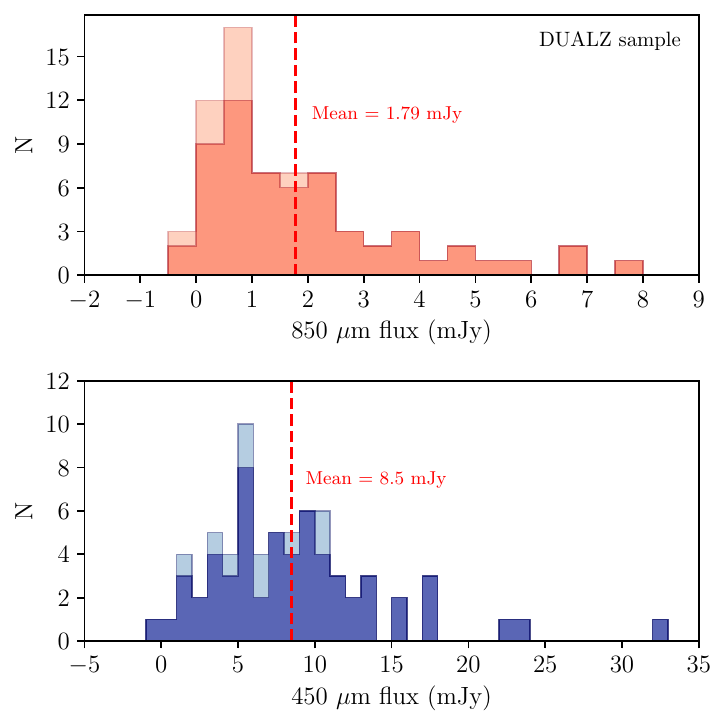}
    \caption{Distribution of 450~$\mu$m and 850~$\mu$m flux densities measured in the SCUBA-2 images at the positions of the 69 DUALZ sources. The darker colors denote the $>5\sigma$ ALMA sources. We show the mean 450~$\mu$m and 850~$\mu$m fluxes for the full DUALZ sample (dotted lines). The minimum rms noise at 850~$\mu$m is 0.26~mJy and at 450~$\mu$m is 2.8~mJy \citep{cowie22}.}
    \label{fig:measured_scuba}
\end{figure}

The resulting distribution of SCUBA-2 450~$\mu$m and 850~$\mu$m fluxes for the ALMA sample is shown in Figure~\ref{fig:measured_scuba}. The mean of the respective distributions is also shown: The mean 850~$\mu$m flux is 1.79~mJy, which is roughly six times the central rms noise of 0.28~mJy, while the mean 450~$\mu$m flux is 8.5~mJy, or three times the central rms of 2.8~mJy.

\begin{figure*}[!t]
    \centering
    \includegraphics[width=0.95\linewidth]{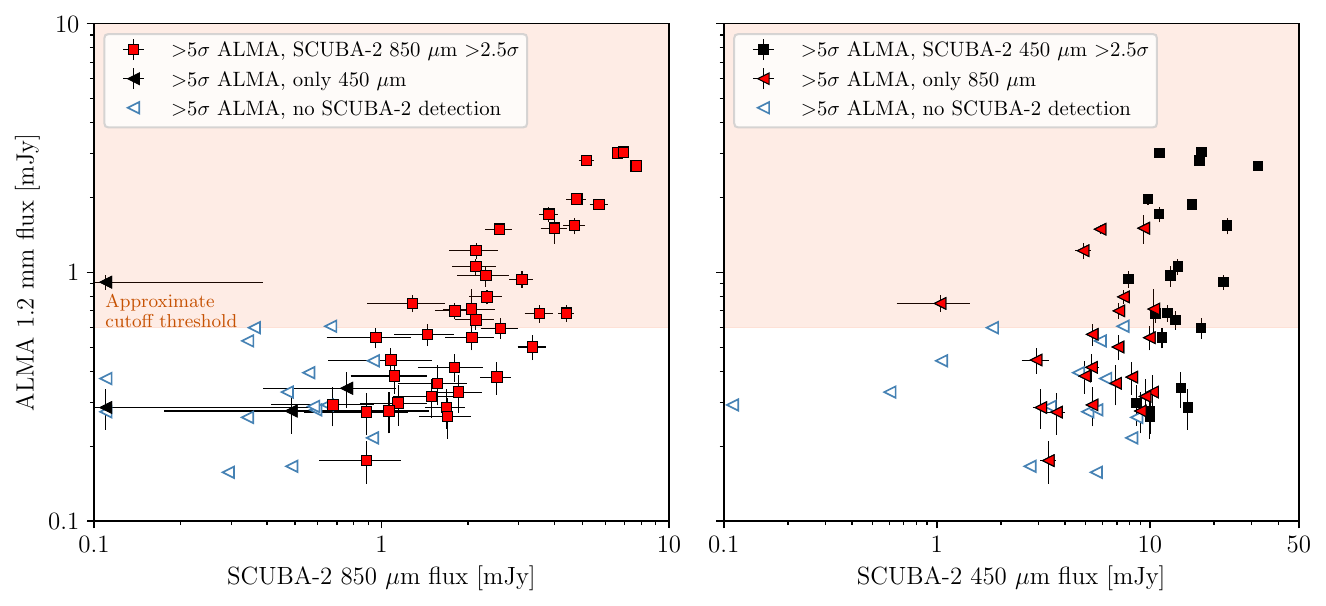}
    \caption{ALMA 1.2~mm flux densities vs. SCUBA-2 850~$\mu$m (left) and 450~$\mu$m (right) flux densities measured at the ALMA positions, restricting to DUALZ sources with $>5\sigma$ ALMA detections. On the left, we distinguish between $>2.5\sigma$ detections at 850~$\mu$m (red squares) and sources detected at $>2.5\sigma$ at 450~$\mu$m but not 850~$\mu$m (black filled triangles); while on the right, we show $>2.5\sigma$ detections at 450~$\mu$m  (black squares) and sources detected at $>2.5\sigma$ at 850~$\mu$m but not 450~$\mu$m (red filled triangles). In both panels, we show sources not detected by SCUBA-2 in either band as blue open triangles, and we show sources with negative measured 850~$\mu$m or 450~$\mu$m SCUBA-2 fluxes at a nominal flux of 0.11~mJy. 
    The shaded region denotes the 1.2~mm flux regime (above a cutoff threshold of $\sim$0.6~mJy) in which the ALMA sources are all detected in the SCUBA-2 data.}
    \label{fig:high_snr_alma}
\end{figure*}

We find that 37 of the ALMA sources (54\%) are detected at $>3\sigma$ at 850~$\mu$m, with an additional six detected at $>2.5\sigma$. All of the direct $>5\sigma$ SCUBA-2 sources previously matched to DUALZ sources are recovered by this method, as are all but three of the $>3\sigma$ SCUBA-2 sources based on the red galaxy priors that were previously matched to the DUALZ sources. Since the $>3\sigma$ SCUBA-2 sources were extracted using the same method as \citet{barger23}, the fact that three of them are at lower significance here is likely because the prior sample is different (ALMA versus JWST) and thus our extraction is deblending the sources differently. Combining the 43 ALMA sources detected at $>2.5\sigma$ in the 850~$\mu$m map with the three that were previously identified as SCUBA-2 counterparts but were not selected by this method, we estimate an upper limit on the fraction of ALMA sources detected at 850~$\mu$m of 46/69 (67\%). By searching for peaks within 4$''$ of 5000 random positions in the SCUBA-2 map, we estimate that the false positive rate for this method is $\sim$8\% at 2.5$\sigma$; in other words, we expect at most four of our $>2.5\sigma$ detections could be false positives.

Of the 69 ALMA sources, 24 are detected at $>2.5\sigma$ at 450~$\mu$m. Eighteen of these are also detected at 850~$\mu$m, with a median $f_{450 \mu{\rm m}}$/$f_{850 \mu{\rm m}}$ ratio of 3.75. By measuring the fluxes and errors at random positions, we estimate that the false positive rate is 3\%. Thus, we expect that at most one of our 24 detections could be a false positive. Of the six sources detected at 450~$\mu$m but not detected at 850~$\mu$m, one source corresponds to one of the three previously-identified red priors that was deblended differently in the 850~$\mu$m extraction, as mentioned above. This leaves five unique sources detected at 450~$\mu$m that are not counted in the 850~$\mu$m total. If we combine the 46 sources detected at 850~$\mu$m with these five sources, then 51/69 (74\%) of the ALMA sources have SCUBA-2 $>2.5\sigma$ detections in at least one band.

In Table~\ref{tab:samples}, we include a summary of the fraction of DUALZ ALMA sources detected at SCUBA-2 850~$\mu$m and/or 450~$\mu$m. We list the measured SCUBA-2 450~$\mu$m and 850~$\mu$m fluxes, errors, and S/N ratios for all the DUALZ sources in Table~\ref{tab:measuredscuba}.

\begin{figure*}[t]
    \centering
    \includegraphics[width=0.98\linewidth]{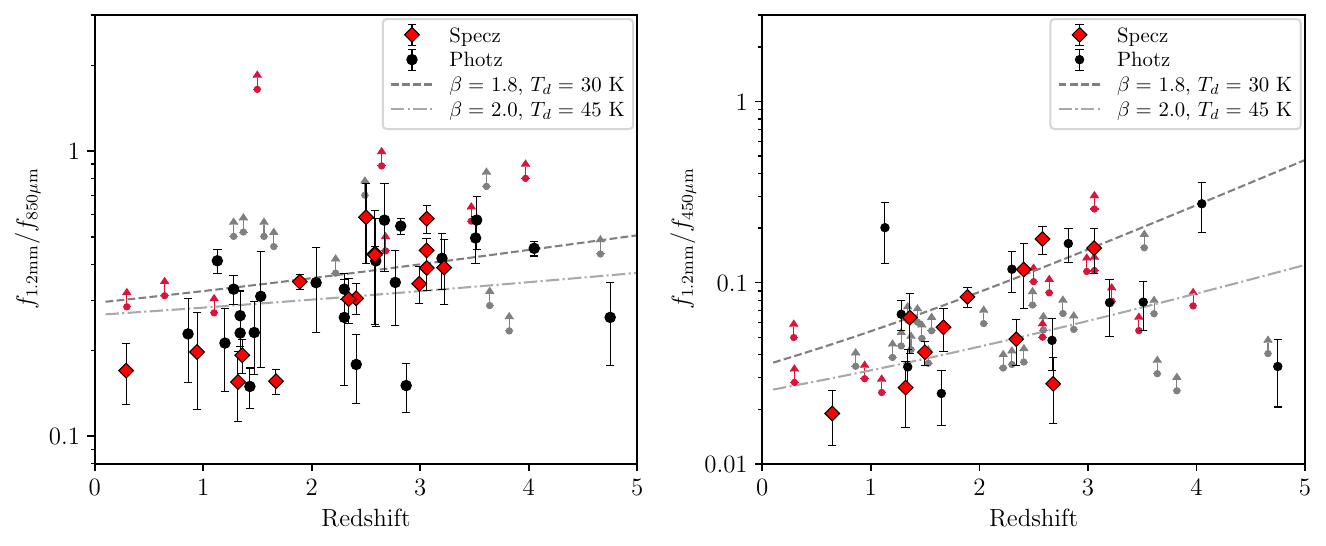}
    \caption{For sources in the DUALZ catalog with ALMA S/N~$>$~5: (left) 1.2~mm to 850~$\mu$m flux ratio vs. redshift and (right) 1.2~mm to 450~$\mu$m flux ratio vs. redshift. In each panel, for sources with $>2.5\sigma$ detections in the relevant SCUBA-2 bands, photzs are shown as black circles and speczs as red diamonds. The 2$\sigma$ lower limits (upward-pointing arrows; speczs are red circles and photzs are gray circles) are shown for non-detections at 850~$\mu$m (left panel) or 450~$\mu$m (right panel).
    In each panel, we plot the expected flux ratios across redshift for modified blackbodies with varying emissivity indices, $\beta$, and dust temperatures, $T_d$ (dotted and dashed lines---see legend).
    }
    \label{fig:ratio_redshift}
\end{figure*}

\section{Discussion}

\subsection{Properties of ALMA Sources Detected in SCUBA-2 versus Non-detections} 

The analysis of the previous section has shown that nearly 75\% of the ALMA sources in DUALZ are also detected in SCUBA-2 at either 850~$\mu$m, 450~$\mu$m, or both. Here we compare the flux and redshift properties of these sources to study any correlations with SCUBA-2 detection rate. 

We have already seen from Figure~\ref{fig:1mm_850um} that most of the DUALZ sources not in the SCUBA-2 catalogs of \citet{barger23} are those with lower ALMA S/N ratios and/or 1.2~mm fluxes. In Figure~\ref{fig:high_snr_alma}, we make this more quantitative by comparing the ALMA fluxes to the SCUBA-2 850~$\mu$m and 450~$\mu$m fluxes directly measured at each ALMA position, limiting the sample to the 59 DUALZ sources with ALMA S/N~$>$~5. 

We first see that down to a 1.2~mm flux limit of $\sim$0.6~mJy (shaded region), we detect all the $>5\sigma$ ALMA sources in at least one SCUBA-2 band at $>2.5\sigma$, but below this limit, we start to lose SCUBA-2 detections. Only one of the ALMA sources (\#23 in Table~\ref{tab:measuredscuba}) above the 0.6~mJy limit at 1.2~mm is detected only at 450~$\mu$m and not at 850~$\mu$m; this source is a low-redshift ($z_{\rm spec}$ = 1.498 from \citealt{munozarancibia+23}) galaxy that is one of the brightest detections at 450~$\mu$m (22.1~mJy). We can also see that the loss of SCUBA-2 detections below the 0.6~mJy cut at 1.2~mm is primarily an issue of SCUBA-2 image depth: The strong correlation between 1.2~mm and 850~$\mu$m flux implies that fainter ALMA sources will start to fall below the SCUBA-2 850~$\mu$m 2.5$\sigma$ detection limit ($\sim$0.8~mJy in the current images). 

We next use the available redshifts for these sources to investigate how the SCUBA-2-detected fraction (i.e., the fraction of ALMA sources with $>2.5\sigma$ 450~$\mu$m and/or 850~$\mu$m detections) varies with redshift. \citet{fujimoto+23b} published photometric redshifts (photzs) estimated with \textsc{prospector} \citep{johnson21} and \textsc{eazy} \citep{brammer08} and spectroscopic redshifts (speczs) from NIRSpec prism observations for the DUALZ catalog (both the speczs and photzs are discussed in more detail in \citealt{wang23}). For the 27 sources with speczs, they find excellent agreement with the photz estimates. 

In Figure~\ref{fig:ratio_redshift}, we show the 1.2~mm to 850~$\mu$m flux ratios (left) and  1.2~mm to 450~$\mu$m flux ratios (right) versus redshift for the $>5\sigma$ DUALZ sources with redshifts (two $>5\sigma$ ALMA sources have neither speczs nor photzs and are not shown). We show 2$\sigma$ lower limits (upward-pointing arrows) in cases where the SCUBA-2 fluxes have less than $2.5\sigma$ significance. 
In each panel, we also show the expected tracks for simple modified blackbody models of the form $f_\nu \propto \nu^\beta B(\nu, T_d)$ with varying emissivity indices, $\beta$, and dust temperatures, $T_d$ (see figure legend; e.g., \citealt{mckay23}). In the right panel, we see good alignment between the measured 1.2~mm to 450~$\mu$m flux ratios and the models; while in the left panel, the measured 1.2~mm to 850~$\mu$m flux ratios show a larger degree of scatter. This is likely because the 850~$\mu$m and 1.2~mm bands lie close to each other and are therefore more susceptible to uncertainties in the flux ratios.  

There is a slight evolution with redshift in the SCUBA-2-detected fraction of ALMA sources. Specifically, the SCUBA-2-detected fraction for the $>5\sigma$ ALMA sources with $1<z<3$ is 81\%, while for $3<z<5$ it is 60\%. From the modified blackbody tracks in both panels of Figure~\ref{fig:ratio_redshift}, we can see that the expected 1.2~mm to 850~$\mu$m and 1.2~mm to 450~$\mu$m flux ratios increase from $z=0$ to $z=5$, which suggests that some of these galaxies may be undetected simply because of the deeper ALMA flux limits combined with the higher mm to submm flux ratios as we move to higher redshift.

Other authors have compared submm/mm flux ratios to published speczs and photzs \citep{casey13,lim20,barger22,cowie23} or used them to make rough redshift estimates for sources whose optical/near-infrared (NIR) counterparts are faint or not identified \citep[e.g.,][]{wang19}. These estimates rely on the fact that the submm bands lie closer to the peak of the dust emission in the FIR and therefore have flux ratios that change substantially as the spectral energy distribution (SED) is redshifted. 

For the DUALZ sources with ALMA S/N~$>$~5, we find that the 1.2~mm to 450~$\mu$m flux ratio (right panel of Figure~\ref{fig:ratio_redshift}) shows a steeper dependence on redshift than the 1.2~mm to 850~$\mu$m ratio, reflecting the larger separation between the observed wavelengths. The best-fit line is $\log(f_{1.2 {\rm mm}}/f_{450 \mu{\rm m}}) = 0.16z-1.50$, with some scatter. The 1.2~mm to 850~$\mu$m flux ratio has a slightly shallower best-fit line of $\log(f_{1.2 {\rm mm}}/f_{450 \mu{\rm m}}) = 0.10z-0.70$. Although other studies \citep[e.g.,][]{barger22} have found a good redshift correlation with the 450~$\mu$m to 850~$\mu$m ratio, there are not enough sources with both 450~$\mu$m and 850~$\mu$m detections to constrain this relationship for the DUALZ sample. Deepening the 450~$\mu$m imaging on the field (which is substantially shallower than the data used in \citealt{barger22}) would help constrain these relationships and could help to further refine the redshifts and dust properties of these galaxies.

We note that nearly all of the DUALZ sources are at $z\lesssim4$, with just two sources between $4.5<z<5$ and only one source at $z>5$. \citet{fujimoto+23b} discuss the latter object (\#66 in Table~\ref{tab:measuredscuba}; this object is not shown in Figure~\ref{fig:ratio_redshift} since it has an ALMA S/N of 4.9), which has $z_{\rm phot} = 9.89^{+1.22}_{-9.20}$, suggesting that it may be an example of a JWST-dark galaxy. We find that this source is not detected in either the 450~$\mu$m or 850~$\mu$m data. This is not unexpected if the source is real and located at $z\sim10$, since a simple modifed blackbody fit to the ALMA point would suggest the 850~$\mu$m flux should be around 0.20~mJy (with some dependence on the assumed $T_d$ and $\beta$), just below the noise level. Nevertheless, the 450~$\mu$m and 850~$\mu$m upper limits could provide additional constraints on the photz fitting.

\subsection{Using Red JWST Priors to Identify Counterparts to Faint SCUBA-2 Galaxies}

The expanded ALMA data in the A2744 field provided by the DUALZ survey allows us to test how well the red NIRCam selection of \citet{barger23} picks out the ALMA counterparts to submillimeter-bright DSFGs. As described in Section~\ref{sec:alma_props}, 31/39 of the red priors with $>3\sigma$ SCUBA-2 fluxes are also in the DUALZ catalog of \citet{fujimoto+23b}, with another six detected at $>2.5\sigma$ in the DUALZ map. This corresponds to a total fraction of red galaxies with $>3\sigma$ SCUBA-2 fluxes confirmed by ALMA of $\sim$95\%. 

This result emphasizes the utility of the red color selection as a method for probing to fainter submillimeter fluxes, and shows that one can obtain large samples of faint DSFGs in fields with both deep SCUBA-2 and JWST coverage. Furthermore, this red selection provides an efficient way to identify sources for high-resolution ALMA follow-up imaging or for millimeter spectral scans at a much lower investment of observing time than large mosaicked surveys \citep[e.g.,][]{cowie22}.

We note that a red NIRCam prior selection used to select the SCUBA-2 $>3\sigma$ sources does introduce a potential selection bias against faint submillimeter/millimeter sources with less red NIR counterparts, including some of the other ALMA sources in the \citet{fujimoto+23b} catalog. In an upcoming paper \citep{mckayinprep}, we will further discuss the selection effects of using red counterparts to pick out DSFGs and study the properties of red sources detected in the submillimeter/millimeter versus non-detections. In addition, we note that upcoming ALMA Band~7 follow-up observations (ALMA project \#2023.1.00468.S; PI: F. Bauer) of over 80 SCUBA-2 850~$\mu$m-selected DSFGs across three lensing cluster fields with JWST data (including A2744) will help to pin down the fraction of SCUBA-2 sources that can be recovered with red priors. Furthermore, these observations will greatly increase the number of faint DSFGs with ALMA detections while constraining the cosmic variance between fields. Finally, the combination of the ALMA and SCUBA-2 data with JWST NIRCam imaging will make it possible to measure stellar masses and star formation rates from panchromatic SED fits.

\section{Summary}
 
We used complementary ALMA and SCUBA-2 observations to compare submillimeter and millimeter galaxy selections in the A2744 cluster field. We showed that 15/17 of the SCUBA-2 $>5\sigma$ sources are detected in the DUALZ ALMA 1.2~mm mosaic of \citet{fujimoto+23b}, with a low multiplicity of 13\%. We also showed that about 95\% of the red galaxy priors identified as $>3\sigma$ SCUBA-2 sources in \citet{barger23} are detected by ALMA. Six of these are not detected outright in the DUALZ catalog but are detected at lower significance ($>2.5\sigma$) in the ALMA image. From the high ALMA detection fraction, we concluded that red prior selection combined with SCUBA-2 data is a straightforward and efficient way to select faint DSFGs. 

We also analyzed the SCUBA-2 450~$\mu$m and 850$\mu$m fluxes of the DUALZ sample and found that 74\% of the ALMA sources are detected in at least one submillimeter band: 46/69 are detected at $>2.5\sigma$ at 850~$\mu$m, with an additional five detected at $>2.5\sigma$ in the 450~$\mu$m data but not at 850~$\mu$m. All of the $>5\sigma$ ALMA sources that are not detected in SCUBA-2 have 1.2~mm fluxes $\lesssim$~0.6~mJy. Thus, we concluded that the lack of detections in SCUBA-2 is primarily an issue of depth.  

Using the photzs and speczs for the DUALZ sample, we compared the variation in the measured submillimeter/millimeter fluxes and the SCUBA-2 detection fraction with redshift. We found that the SCUBA-2 detection fraction of ALMA sources dropped slightly beyond $z=3$, possibly due to higher millimeter to submillimeter flux ratios combined with the deeper ALMA flux limits. Furthermore, we showed that the 1.2~mm to 450~$\mu$m flux ratio is well correlated with redshift for this sample. Deeper 450~$\mu$m imaging, in particular, could help constrain the redshifts and dust properties of the full ALMA sample.

\vspace{6cm}

\begin{acknowledgements}
{
We thank the anonymous referee for constructive suggestions that helped us to improve the manuscript. We gratefully acknowledge support for this research from 
the William F. Vilas Estate (S.J.M.),
a WARF Named Professorship from the 
University of Wisconsin-Madison Office of the 
Vice Chancellor for Research and Graduate Education with funding from the Wisconsin Alumni Research Foundation (A.J.B.), 
and NASA grant 80NSSC22K0483 (L.L.C.).

The National Radio Astronomy Observatory is a facility of the National Science
Foundation operated under cooperative agreement by Associated Universities, Inc.
This paper makes use of the following ALMA data: 
ADS/JAO.ALMA\#2013.1.00999.S. \\
ADS/JAO.ALMA\#2015.1.01425.S, \\
ADS/JAO.ALMA\#2017.1.01219.S, \\
ADS/JAO.ALMA\#2018.1.00035.L, \\ 
and ADS/JAO.ALMA\#2022.1.00073.S. \\ 
ALMA is a partnership of ESO (representing its member states), NSF (USA), and NINS (Japan), 
together with NRC (Canada), MOST and ASIAA (Taiwan), and KASI (Republic of Korea),
 in cooperation with the Republic of Chile. The Joint ALMA Observatory is operated by 
 ESO, AUI/NRAO, and NAOJ.
 
The James Clerk Maxwell Telescope is operated by the East Asian Observatory on 
behalf of The National Astronomical Observatory of Japan, Academia Sinica Institute 
of Astronomy and Astrophysics, the Korea Astronomy and Space Science Institute, 
the National Astronomical Observatories of China and the Chinese Academy of 
Sciences (grant No.~XDB09000000), with additional funding support from the Science 
and Technology Facilities Council of the United Kingdom and participating universities 
in the United Kingdom and Canada. 

We wish to recognize and acknowledge 
the very significant cultural role and reverence that the summit of Maunakea has always 
held within the indigenous Hawaiian community. We are most fortunate to have the 
opportunity to conduct observations from this mountain.
}
\end{acknowledgements}

\facilities{ALMA, JCMT}
\software{astropy \citep{astropy:2022}}

\bibliography{bib}{}
\bibliographystyle{aasjournal}


\begin{deluxetable*}{lcccc|lcc}
\tabletypesize{\scriptsize }
\tablewidth{\textwidth}
\tablecaption{\label{tab:samples}
Summary of SCUBA-2 and ALMA Samples and Detection Fractions}
\tablehead{
\multicolumn{5}{c}{SCUBA-2 \citep{barger23}} &  \multicolumn{3}{c}{DUALZ \citep{fujimoto+23b}}\\[1mm]
& \multicolumn{2}{c}{$> 5 \sigma$ (direct)} &  \multicolumn{2}{c}{$> 3 \sigma$ (red priors)}  & & & \\
& \colhead{Number} & \colhead{\%} & \colhead{Number} & \colhead{\%}  & & \colhead{Number} & \colhead{\%}}
\startdata
Total in ALMA coverage & 17 & 100\% & 39 & 100\% & Total & 69& 100\%\\
With red counterparts & 15\tablenotemark{\scriptsize a}\tablenotemark{\scriptsize b}& 88\% & 39 & 100\%  & $>5\sigma$ in ALMA maps& 59& 86\%\\
With DUALZ counterparts & 15\tablenotemark{\scriptsize a}& 88\% & 31 & 80\% &  $>2.5\sigma$, SCUBA-2 850~$\mu$m& 43\tablenotemark{\scriptsize c}& 62\%\\
{\bf $>2.5\sigma$ ALMA detections} & {\bf 15}\tablenotemark{\scriptsize a}& {\bf 88\%} & {\bf 37} & {\bf 95\% }&  $>2.5\sigma$, SCUBA-2 450~$\mu$m& 24& 35\%\\
&&&& &  $>2.5\sigma$, both SCUBA-2 bands& 18& 26\%\\
&&&& &  $>2.5\sigma$, at least one SCUBA-2 band& 49& 71\%\\
&&&& &  {\bf Total with SCUBA-2 detections} & {\bf 51\tablenotemark{\scriptsize c}} & {\bf 74\%}\\
&&&& &&&\\
\enddata
\tablecomments{Thirteen of the $>5\sigma$ direct SCUBA-2 detections correspond to 12 of the $>3\sigma$ red-prior-selected SCUBA-2 sources.}
\tablenotetext{\scriptsize a}{Two SCUBA-2 sources match to the same counterpart (see Section~3).}
\tablenotetext{\scriptsize b}{One red counterpart was corrected by the ALMA data to a different red prior slightly further from the SCUBA-2 centroid (see Section~3).}
\tablenotetext{\scriptsize c}{Three ALMA sources which correspond to red-prior-selected $>3\sigma$ SCUBA-2 850~$\mu$m sources were not detected with SCUBA-2 at 850~$\mu$m using the ALMA priors; however, one was detected at 450~$\mu$m (see Section~4). We include these sources in the final total.}

\end{deluxetable*}

\vskip -1cm

\startlongtable
\begin{deluxetable*}{cccccccccccc}
\tabletypesize{\scriptsize}
\tablewidth{\textwidth}
\tablecaption{\label{tab:measuredscuba}
SCUBA-2 450~$\mu$m and 850~$\mu$m Properties of DUALZ Catalog}
\tablehead{
\colhead{} & \multicolumn{2}{c}{ALMA}&\multicolumn{3}{c}{SCUBA-2}
&\multicolumn{3}{c}{SCUBA-2}& \multicolumn{3}{c}{ALMA}\\
\colhead{No.} & \colhead{R.A.}&\colhead{Decl.}
&\colhead{$f_{450\,\mu{\rm m}}$} & \colhead{Error} &\colhead{S/N} & \colhead{$f_{850\,\mu{\rm m}}$} & \colhead{Error} &\colhead{S/N} &\colhead{$f_{\rm 1.2\,mm}$} & \colhead{Error} &\colhead{S/N}\\
    & \colhead{J2000.0} & \colhead{J2000.0} & \multicolumn{2}{c}{(mJy)}& & \multicolumn{2}{c}{(mJy)} & & \multicolumn{2}{c}{(mJy)}&}

\colnumbers

\startdata
1 & 3.6036340 & -30.429562 & 3.65 & 3.82 & 1.0 & 0.89 & 0.35 & 2.5 & 0.27 & 0.05 & 5.3 \\
2 & 3.5755829 & -30.424378 & 11.1 & 3.39 & 3.3 & 6.62 & 0.30 & 22.2 & 3.02 & 0.12 & 41.4 \\
3 & 3.5863515 & -30.425044 & 7.52 & 3.46 & 2.2 & 2.33 & 0.31 & 7.6 & 0.80 & 0.05 & 15.0 \\
4 & 3.6056155 & -30.418061 & 10.0 & 3.49 & 2.9 & 0.49 & 0.31 & 1.6 & 0.28 & 0.05 & 5.3 \\
5 & 3.5760482 & -30.413200 & 17.5 & 3.00 & 5.8 & 6.96 & 0.27 & 25.9 & 3.04 & 0.12 & 45.2 \\
6 & 3.6085748 & -30.414591 & 11.4 & 3.46 & 3.3 & 0.96 & 0.31 & 3.1 & 0.55 & 0.05 & 10.5 \\
7 & 3.5650518 & -30.412200 & 5.11 & 3.07 & 1.7 & -0.14 & 0.27 & -0.5 & 0.28 & 0.05 & 6.1 \\
8 & 3.5690146 & -30.402798 & 5.36 & 2.93 & 1.8 & 0.68 & 0.26 & 2.6 & 0.29 & 0.05 & 6.3 \\
9 & 3.5723726 & -30.395954 & 4.63 & 2.93 & 1.6 & 0.56 & 0.26 & 2.1 & 0.40 & 0.05 & 7.5 \\
10 & 3.6276995 & -30.394278 & 13.5 & 3.92 & 3.4 & 2.13 & 0.37 & 5.8 & 1.06 & 0.08 & 21.5 \\
11 & 3.6005232 & -30.396158 & 3.34 & 2.97 & 1.1 & 0.89 & 0.28 & 3.2 & 0.18 & 0.03 & 5.2 \\
12 & 3.6175369 & -30.395013 & 5.63 & 3.45 & 1.6 & 0.59 & 0.32 & 1.8 & 0.28 & 0.06 & 6.0 \\
13 & 3.5667961 & -30.394888 & 7.16 & 3.00 & 2.4 & 1.80 & 0.27 & 6.8 & 0.70 & 0.05 & 15.3 \\
14 & 3.6326635 & -30.393642 & 7.47 & 4.08 & 1.8 & 0.67 & 0.38 & 1.8 & 0.61 & 0.06 & 14.0 \\
15 & 3.6214798 & -30.393109 & 5.36 & 3.57 & 1.5 & 1.44 & 0.34 & 4.3 & 0.56 & 0.06 & 13.2 \\
16 & 3.5809797 & -30.390749 & 3.08 & 2.87 & 1.1 & 1.68 & 0.27 & 6.3 & 0.29 & 0.05 & 5.5 \\
17 & 3.5474111 & -30.388289 & 17.1 & 3.64 & 4.7 & 5.16 & 0.31 & 16.8 & 2.82 & 0.07 & 55.2 \\
18 & 3.5327226 & -30.387320 & 6.20 & 4.38 & 1.4 & 0.08 & 0.36 & 0.2 & 0.37 & 0.06 & 6.6 \\
19 & 3.5719637 & -30.383018 & 12.1 & 3.05 & 4.0 & 4.40 & 0.28 & 15.7 & 0.68 & 0.05 & 13.5 \\
20 & 3.5190588 & -30.386326 & 9.36 & 5.10 & 1.8 & 3.99 & 0.41 & 9.6 & 1.50 & 0.20 & 8.9 \\
21 & 3.5825067 & -30.385468 & 5.85 & 2.92 & 2.0 & 2.58 & 0.28 & 9.3 & 1.49 & 0.06 & 42.0 \\
22 & 3.5850010 & -30.381794 & 11.1 & 2.98 & 3.7 & 3.82 & 0.29 & 13.2 & 1.71 & 0.12 & 24.5 \\
23 & 3.5732507 & -30.383497 & 22.1 & 3.01 & 7.3 & 0.05 & 0.28 & 0.2 & 0.91 & 0.06 & 24.6 \\
24 & 3.6357675 & -30.382387 & 1.82 & 3.98 & 0.5 & 0.36 & 0.43 & 0.8 & 0.60 & 0.06 & 11.3 \\
25 & 3.5126766 & -30.380966 & 10.4 & 5.27 & 2.0 & 2.06 & 0.42 & 4.9 & 0.71 & 0.15 & 6.9 \\
26 & 3.5920643 & -30.380487 & 5.85 & 3.02 & 1.9 & 0.34 & 0.30 & 1.1 & 0.53 & 0.06 & 11.5 \\
27 & 3.5812925 & -30.380250 & 0.60 & 3.04 & 0.2 & 0.47 & 0.29 & 1.6 & 0.33 & 0.05 & 9.1 \\
28 & 3.5797072 & -30.378413 & 7.94 & 3.04 & 2.6 & 3.08 & 0.30 & 10.4 & 0.94 & 0.07 & 21.5 \\
29 & 3.5599876 & -30.377804 & 3.40 & 3.35 & 1.0 & 0.58 & 0.30 & 1.9 & 0.29 & 0.06 & 5.9 \\
30 & 3.5579207 & -30.377236 & 8.23 & 3.43 & 2.4 & 2.52 & 0.31 & 8.1 & 0.38 & 0.06 & 6.7 \\
31 & 3.5823806 & -30.377170 & 8.65 & 3.08 & 2.8 & 1.14 & 0.30 & 3.8 & 0.30 & 0.06 & 5.8 \\
32 & 3.5995954 & -30.374707 & 4.95 & 3.22 & 1.5 & 1.11 & 0.33 & 3.4 & 0.38 & 0.06 & 7.4 \\
33 & 3.5103148 & -30.375438 & 6.91 & 5.18 & 1.3 & 1.57 & 0.42 & 3.7 & 0.36 & 0.07 & 6.1 \\
34 & 3.5543533 & -30.371955 & 13.2 & 3.60 & 3.7 & 2.13 & 0.33 & 6.4 & 0.64 & 0.06 & 13.1 \\
35 & 3.6172782 & -30.368795 & 9.80 & 3.58 & 2.7 & 4.78 & 0.38 & 12.5 & 1.97 & 0.10 & 35.6 \\
36 & 3.5237288 & -30.371469 & 10.3 & 4.53 & 2.3 & 1.85 & 0.38 & 4.8 & 0.33 & 0.06 & 8.7 \\
37 & 3.5638035 & -30.367614 & 9.99 & 3.54 & 2.8 & 1.70 & 0.34 & 4.9 & 0.26 & 0.05 & 5.4 \\
38 & 3.5375675 & -30.365649 & 14.0 & 4.06 & 3.4 & 0.76 & 0.37 & 2.0 & 0.34 & 0.06 & 6.9 \\
39 & 3.5311608 & -30.361294 & 7.09 & 4.14 & 1.7 & 3.35 & 0.38 & 8.8 & 0.50 & 0.06 & 9.2 \\
40 & 3.6006013 & -30.362714 & 1.04 & 3.70 & 0.3 & 1.28 & 0.39 & 3.3 & 0.75 & 0.06 & 16.0 \\
41 & 3.5391400 & -30.360300 & 17.4 & 3.97 & 4.4 & 2.59 & 0.38 & 6.8 & 0.60 & 0.06 & 10.6 \\
42 & 3.5362850 & -30.360378 & 32.1 & 4.02 & 8.0 & 7.69 & 0.38 & 20.2 & 2.68 & 0.09 & 50.4 \\
43 & 3.5211455 & -30.360680 & 9.95 & 4.40 & 2.3 & 2.06 & 0.40 & 5.2 & 0.55 & 0.06 & 11.1 \\
44 & 3.5990300 & -30.359757 & 10.6 & 3.78 & 2.8 & 3.55 & 0.40 & 8.9 & 0.68 & 0.06 & 12.2 \\
45 & 3.5938354 & -30.356617 & 23.0 & 3.98 & 5.8 & 4.70 & 0.42 & 11.2 & 1.54 & 0.12 & 24.7 \\
46 & 3.5237434 & -30.359221 & -0.51 & 4.33 & -0.1 & 0.65 & 0.39 & 1.7 & 0.29 & 0.06 & 5.9 \\
47 & 3.5926433 & -30.356140 & 2.93 & 4.05 & 0.7 & 1.08 & 0.43 & 2.5 & 0.44 & 0.05 & 8.3 \\
48 & 3.5583154 & -30.354969 & 4.86 & 3.92 & 1.2 & 2.13 & 0.42 & 5.1 & 1.22 & 0.09 & 23.8 \\
49 & 3.5463709 & -30.353476 & 9.03 & 3.92 & 2.3 & 1.06 & 0.40 & 2.6 & 0.28 & 0.05 & 6.8 \\
50 & 3.5491494 & -30.352245 & 15.8 & 3.93 & 4.0 & 5.71 & 0.41 & 13.9 & 1.87 & 0.07 & 39.3 \\
51 & 3.5305552 & -30.352317 & 9.51 & 4.10 & 2.3 & 1.49 & 0.40 & 3.8 & 0.32 & 0.06 & 7.6 \\
52 & 3.5612866 & -30.326963 & 8.22 & 4.27 & 1.9 & 0.93 & 0.46 & 2.0 & 0.22 & 0.04 & 5.4 \\
53 & 3.5555550 & -30.335627 & 8.67 & 4.15 & 2.1 & 0.34 & 0.45 & 0.8 & 0.26 & 0.05 & 6.3 \\
54 & 3.5528434 & -30.344173 & 1.05 & 4.07 & 0.3 & 0.94 & 0.44 & 2.1 & 0.44 & 0.05 & 8.8 \\
55 & 3.5641531 & -30.344466 & 12.5 & 4.21 & 3.0 & 2.31 & 0.47 & 4.9 & 0.97 & 0.09 & 17.1 \\
56 & 3.5611003 & -30.330043 & 5.33 & 4.22 & 1.3 & 1.80 & 0.46 & 3.9 & 0.42 & 0.05 & 8.8 \\
57 & 3.5566688 & -30.384299 & 10.3 & 3.31 & 3.1 & 0.66 & 0.29 & 2.3 & 0.26 & 0.06 & 4.5 \\
58 & 3.6384449 & -30.381213 & 6.38 & 4.07 & 1.6 & 1.51 & 0.45 & 3.4 & 0.31 & 0.07 & 4.6 \\
59 & 3.5582541 & -30.374437 & 5.58 & 3.47 & 1.6 & 0.7 & 0.32 & 2.2 & 0.26 & 0.06 & 4.7 \\
60 & 3.5923715 & -30.368396 & 4.90 & 3.42 & 1.4 & 0.91 & 0.35 & 2.6 & 0.26 & 0.06 & 4.6 \\
61 & 3.5580404 & -30.331104 & 10.6 & 4.20 & 2.5 & 0.16 & 0.46 & 0.3 & 0.26 & 0.05 & 4.9 \\
62 & 3.5456886 & -30.337678 & 8.42 & 4.1 & 2.1 & 0.66 & 0.44 & 1.5 & 0.24 & 0.05 & 4.5 \\
63 & 3.5671705 & -30.341419 & 5.84 & 4.32 & 1.4 & 0.07 & 0.48 & 0.1 & 0.24 & 0.05 & 4.4 \\
64 & 3.5618008 & -30.330958 & 15.1 & 4.22 & 3.6 & -0.06 & 0.46 & -0.1 & 0.29 & 0.05 & 5.4 \\
65 & 3.5685704 & -30.329642 & 3.05 & 4.32 & 0.7 & -0.13 & 0.46 & -0.3 & 1.14 & 0.24 & 4.7 \\
66 & 3.5938125 & -30.408263 & 1.16 & 3.02 & 0.4 & 0.47 & 0.27 & 1.7 & 0.15 & 0.05 & 4.9 \\
67 & 3.6017503 & -30.407851 & 5.60 & 3.17 & 1.8 & 0.29 & 0.29 & 1.0 & 0.16 & 0.04 & 5.2 \\
68 & 3.5932584 & -30.384375 & 2.74 & 2.95 & 0.9 & 0.49 & 0.29 & 1.7 & 0.17 & 0.05 & 5.0 \\
69 & 3.5902747 & -30.400439 & 6.64 & 2.89 & 2.3 & 0.76 & 0.27 & 2.9 & 0.18 & 0.04 & 4.2 \\
\enddata

\tablecomments{Fluxes have not been corrected for magnification.}
\end{deluxetable*}

\end{document}